\documentclass{aa}
\bibpunct{(}{)}{;}{a}{}{,}
\usepackage{graphicx}
\usepackage{txfonts}
\usepackage{hyperref}
\usepackage{txfonts}
\usepackage{colortbl}
\usepackage{orcidlink}
\usepackage{lscape}
\usepackage{amsfonts,amsmath,natbib,xcolor,subfigure,gensymb,longtable,ulem}
\usepackage{hyperref}
\usepackage{siunitx}
\hypersetup{breaklinks,colorlinks,citecolor=blue,linkcolor=blue,urlcolor=red}
\defcitealias{Liang09}{L09}

\begin{document}

\title{GRB X-ray plateaus as evidence that the afterglow begins before the prompt gamma-ray emission}

\subtitle{}
    
\titlerunning{GRB X-ray plateaus and the true start of afterglows}

\author{
   C.~Guidorzi\thanks{\texttt{guidorzi[at]fe.infn.it}}\inst{\ref{unife},\ref{infnfe},\ref{inafbo}}\orcidlink{0000-0001-6869-0835}
   \and R.~Maccary\inst{\ref{unife},\ref{inafbo}}\orcidlink{0000-0002-8799-2510}
   \and M.~Maistrello\inst{\ref{unife},\ref{inafbo}}\orcidlink{0009-0000-4422-4151}
   \and S.~Kobayashi\inst{\ref{ljmu}}\orcidlink{0000-0001-7946-4200}
   \and M.~Bulla\inst{\ref{unife},\ref{infnfe},\ref{inafte}}\orcidlink{0000-0002-8255-5127}
   \and F.~Frontera\inst{\ref{unife},\ref{inafbo}}\orcidlink{0000-0003-2284-571X}
  }
            
\institute{Department of Physics and Earth Science, University of Ferrara, Via Saragat 1, I-44122 Ferrara, Italy\label{unife}
   \and  INFN –- Sezione di Ferrara, Via Saragat 1, 44122 Ferrara, Italy\label{infnfe}
   \and INAF –- Osservatorio di Astrofisica e Scienza dello Spazio di Bologna, Via Piero Gobetti 101, 40129 Bologna, Italy\label{inafbo}
   \and Astrophysics Research Institute, Liverpool John Moores University, Liverpool Science Park IC2, 146 Brownlow Hill, Liverpool,  L3 5RF, UK \label{ljmu}
   \and INAF, Osservatorio Astronomico d'Abruzzo, via Mentore Maggini snc, 64100 Teramo, Italy \label{inafte}
}
    
\date{Received date / Accepted date }
    
\abstract
% Context
{Most gamma-ray burst (GRB) X-ray afterglow light curves are characterised by a plateau, followed by a normal power-law decay, which is interpreted as afterglow emission, that is radiation emitted by the shocked interstellar medium that is swept up by the blast wave. Despite the numerous alternative interpretations, the origin of the plateau remains unclear. In the early years of the {\it Neil Gehrels Swift Observatory}, it was suggested that the plateau might be afterglow radiation, that started before the prompt gamma-ray emission, and its time profile would be an artefact of assuming the start time of the prompt gamma-ray emission as zero time (the so-called ``prior activity model'').}
% Aims
{We aim to test the plausibility of the prior activity model by leveraging the current {\it Swift} sample of early X-ray afterglows of GRBs with measured redshifts, which is more than eight times larger than the one originally used (463 vs. 56).}
% Methods
{We modelled the GRB rest-frame X-ray afterglow luminosities assuming a simple power-law with the true reference time preceding the prompt gamma-ray emission trigger time by $T_0$ and the X-ray luminosity $L_0$ at the trigger time as free parameters. We tested each case applying both $\chi^2$ and runs tests.}
% Results
{For 90\% GRBs of our sample, the model provided a successful description. In ten cases the afterglow peak is identified and modelled appropriately. Using the 300 GRBs with accurate parameters' estimates, we confirm the anti-correlation between $L_0$ and $T_0$ with $0.7$~dex scatter. In addition, selecting the subsample of 180 from the literature with reliable estimates of isotropic-equivalent released energy $E_{\gamma,{\rm iso}}$, peak luminosity $L_{\gamma,{\rm iso}}$, and intrinsic peak energy $E_{\rm p,i}$ of the $\nu\,F_\nu$ spectrum of the prompt gamma-ray emission, we find a correlation between $L_0$, $T_0$, and $E_{\gamma,{\rm iso}}$ ($0.4$~dex scatter) over nine decades in $L_0$ and common to all kinds of GRBs.}
% Conclusions
{The afterglow likely begins in most cases before the start of the detected prompt gamma-ray emission by a lognormally-distributed rest-frame delay with a mean of $10^3$~s and $0.8$~dex dispersion. As also suggested by the recent discoveries of {\it Einstein Probe} of X-ray emission starting long before the prompt gamma-rays, our results suggest that the occurrence of prior activity could be much more frequent than what has tacitly been assumed so far.}

\maketitle

%%%%%%%%%%%%%%%%%%%%%%%%%%%%%%%%%%%%%%%%%%%%%%%%%%%%
\section{Introduction}
\label{sec:intro}
%%%%%%%%%%%%%%%%%%%%%%%%%%%%%%%%%%%%%%%%%%%%%%%%%%%%
The unexpected discovery of X-ray plateaus, which characterise the majority of the gamma-ray burst (GRB) early X-ray afterglows and led to the definition of ``canonical'' behaviour \citep{Nousek06,Evans09,Margutti13}, prompted a remarkable number of different interpretations. Some of them invoke (1) continuous energy injection into the fireball \citep{FanPiran06,Granot06,Zhang06,Ghisellini07,Stratta18}, which in most cases requires the GRB inner engine to last much longer than the prompt gamma-ray emission; (2) reverse-shock (RS) emission \citep{Genet07,UhmBeloborodov07,Hascoet14b}; (3) rebrightening caused by inhomogeneities in the medium \citep{Toma06}; (4) high-latitude emission from structured jets \citep{Ascenzi20,Oganesyan20,Beniamini20b}; (5) echo from dust scattering of the X-ray prompt emission \citep{Shao08}; (6) forward shock (FS) emission from relativistic ejecta in the coasting phase as they sweep up a wind-like medium \citep{ShenMatzner12,DereliBegue22}.

As a possible way to solve the unreasonably high (for the internal shock model) gamma-ray efficiency, \citet{Ioka06} proposed a so-called ``prior activity model'': the relativistic explosion would begin much earlier ($10^3$--$10^6$~s) than the prompt gamma-ray emission, such that at the time of the prompt gamma-ray emission, an X-ray afterglow emission may have already started. In other words, the zero time of the afterglow would be comparably earlier than the prompt gamma-ray emission. A different zero time value affects the power-law (PL) that models a given data set and we hereafter refer to this as the zero time effect. The possibility of prior activity was further considered by \citet{Yamazaki09}, who noted that the transition from plateau to normal decay that is observed in the canonical X-ray afterglow, could be an artefact of the wrong zero time: by moving backward by $T_0$ ($\sim 10^3$--$10^4$~s) the zero time from the prompt gamma-ray emission trigger time, the plateau-to-normal decay becomes a simple PL (SPL). This scenario naturally accounts for the lack of spectral evolution through the transition. He consequently modelled the whole X-ray afterglow as the superposition of two components: (i) the prompt X-ray counterpart to the gamma-rays, due to internal dissipation, which manifests itself through the initial steep decay and X-ray flares \citep{Falcone07,Chincarini07,Chincarini10}; (ii) the plateau and normal decay as the afterglow which would have begun before the prompt gamma-ray emission and whose reference time is given by the correct zero time.

The possibility that X-ray plateaus might be an artefact of the zero time effect was systematically investigated by \citet[hereafter L09]{Liang09}, who studied the canonical afterglows among the first $\sim400$~GRBs detected by the {\it Neil Gehrels Swift Observatory} \citep{Gehrels04} and compared with the 19 cases that exhibited a SPL from early on all the way to late times. Both SPL and canonical groups were found to share the same prompt gamma-ray emission properties. The PL indices of the canonical afterglows, once they are modelled with the estimated zero time, are comparable with the indices of the SPL sample, thus lending support to the $T_0$ effect interpretation proposed by \citet{Yamazaki09}. The $T_0$ scenario does not imply that emission should be observed since the zero time: the kinetic energy of the ejecta may not necessarily be converted into the internal energy so much in the early phase \citep{Ioka06}.

The sample of GRBs with measured redshift that was available to \citetalias{Liang09} included 56 GRBs. At the time of writing, there are 463 GRBs with measured redshift that triggered the {\it Swift} Burst Alert Telescope (BAT; \citealt{Barthelmy05}) and whose early X-ray afterglow was promptly covered by the {\it Swift} X-Ray Telescope (XRT; \citealt{Burrows05}). We restrict our analysis to GRBs with redshift to measure and compare the intrinsic zero time delays among different events. By leveraging a nearly tenfold richer sample than that used by \citetalias{Liang09}, we aim to test the zero time scenario, extract the distributions of $T_0$ and of the X-ray afterglow luminosity, $L_0$, at the time of the prompt gamma-ray emission and search for correlations with other key observables. Testing the possibility that for most GRBs the initial explosion begins much earlier than the observed prompt gamma-rays, sounds particularly timely in light of the recent discoveries by {\it Einstein Probe} (EP; \citealt{EinsteinProbeMission}) of X-ray emission preceding the harder prompt emission by several hundred seconds in some cases.
Section~\ref{sec:analysis} reports the data selection and analysis, whose results are presented in Section~\ref{sec:results} and discussed in Section~\ref{sec:disc_conc}. The $\Lambda$CDM cosmological parameters of \citet{cosmoPlanck20} were used.

%%%%%%%%%%%%%%%%%%%%%%%%%%%%%%%%%%%%%%%%%%%%%%%%%%%%
\section{Sample selection and data analysis}
\label{sec:analysis}
%%%%%%%%%%%%%%%%%%%%%%%%%%%%%%%%%%%%%%%%%%%%%%%%%%%%
We selected all the GRBs with spectroscopic redshift from January 2005 to July 2, 2025, that were promptly observed with XRT and with at least 5 points in the light curve (LC). For each GRB we obtained the flux LC in the $0.3$--$10$~keV passband from the Leicester repository\footnote{\url{https://www.swift.ac.uk/xrt_curves/}.}. Two corrections were applied to the flux LC: (i) $k$-correction, to remove the bias of different rest-frame energy bands, by dividing the measured flux by $(1+z)^{2-\Gamma}$, where $z$ and $\Gamma$ are redshift and weighted photon index of the average spectrum reported in the spectral catalogue\footnote{\url{https://www.swift.ac.uk/xrt_live_cat/}; whenever both Window Timing (WT) and Photon Counting (PC) mode average spectra were available, we used the weighted average photon index.} (ii) we corrected the flux for photoelectric absorption, by applying the factor between unabsorbed and absorbed flux reported in the catalogue. Finally, each 1--10~keV rest-frame band, unabsorbed flux $f_x$ was converted to luminosity $L_x = 4\pi D_L^2\,f_x$, where $D_L$ is the luminosity distance, and its evolution $L_x(t)$ was expressed as a function of the rest frame time $t = t_{\rm obs}/(1+z)$, where $t_{\rm obs}$ is the observed time since the BAT trigger.

We initially collected 463 GRBs, which included 413 long (L-GRBs), 34 short (S-GRBs), and 16 short bursts with extended emission (SEE-GRBs).\footnote{In the SEE-GRB group we included the long lasting compact object merger candidates, like 060614, 191019A, 211221A.} We modelled each individual LC with Eq.\eqref{eq:mod},
\begin{equation}
    L_x(t)\ =\ L_0\,\Big(1 + \frac{t}{T_0}\Big)^{-\alpha}\;,
    \label{eq:mod}
\end{equation}
which is equivalent to a simple PL, with time origin that is offset back in time by $T_0$, so that
\begin{equation}
    L_x(t')\ =\ L_0\,\Big(\frac{t'}{T_0}\Big)^{-\alpha}\;,
    \label{eq:mod2}
\end{equation}
where $t' = t + T_0$ represents the time measured since the true beginning of the event. $L_0$ is the afterglow luminosity in erg~s$^{-1}$ units at the trigger time $t=0$. $T_0$ is expressed in seconds. In practice, given the PL nature of the model and the fact that both parameters $L_0$ and $T_0$ vary over several decades, we used their logarithms as free parameters\footnote{In such cases, uncertainties on logarithmic quantities are more symmetric, hence more convenient to use.}, and worked on logarithmic fluxes and times.

All the time intervals featuring internal activity in the form of either X-ray flares, X-ray prompt emission or its end that is signalled by the steep decay, were ignored.

\begin{figure}[!h]
   \includegraphics[width=0.45\textwidth]{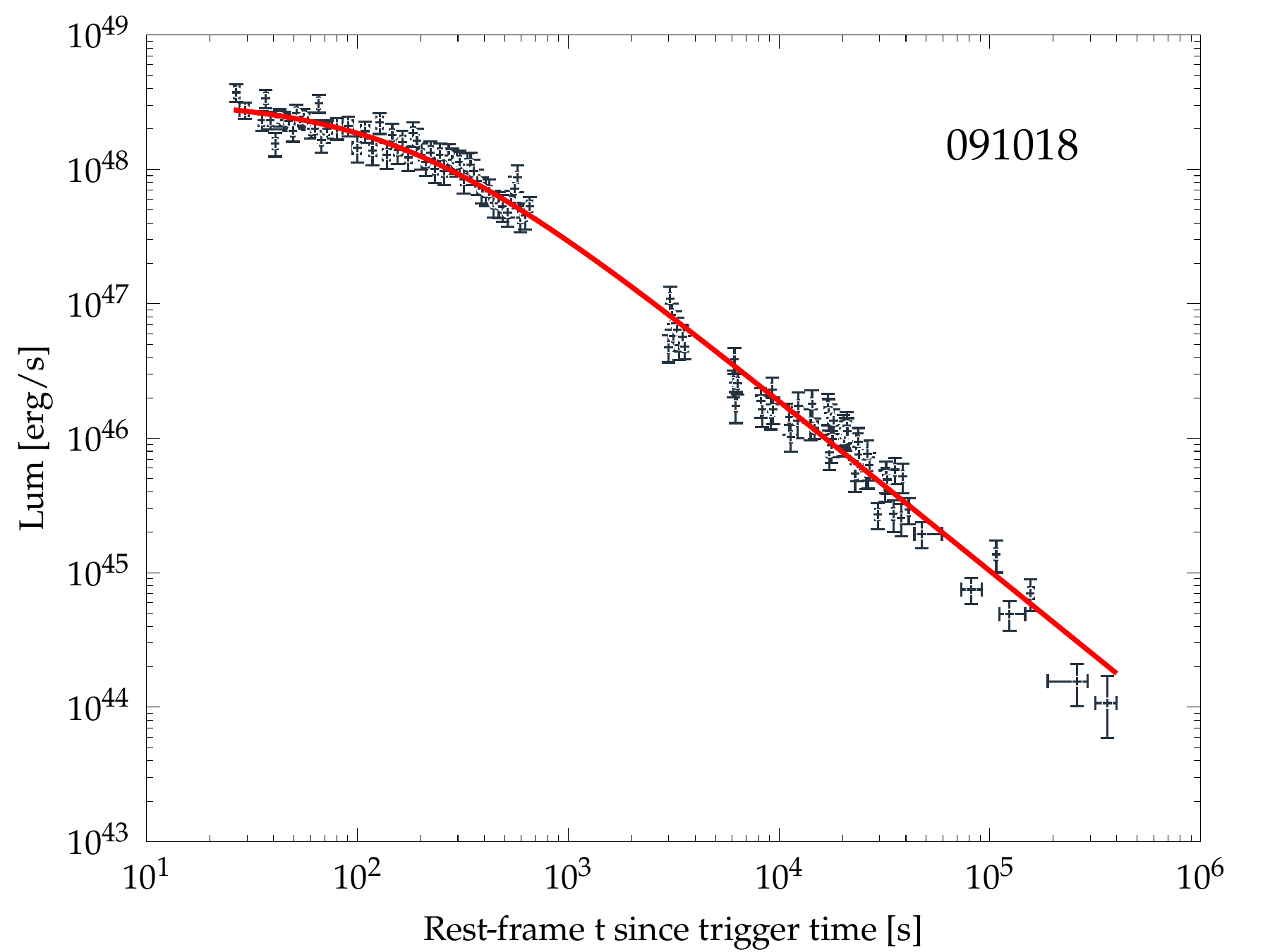}
   \includegraphics[width=0.45\textwidth]{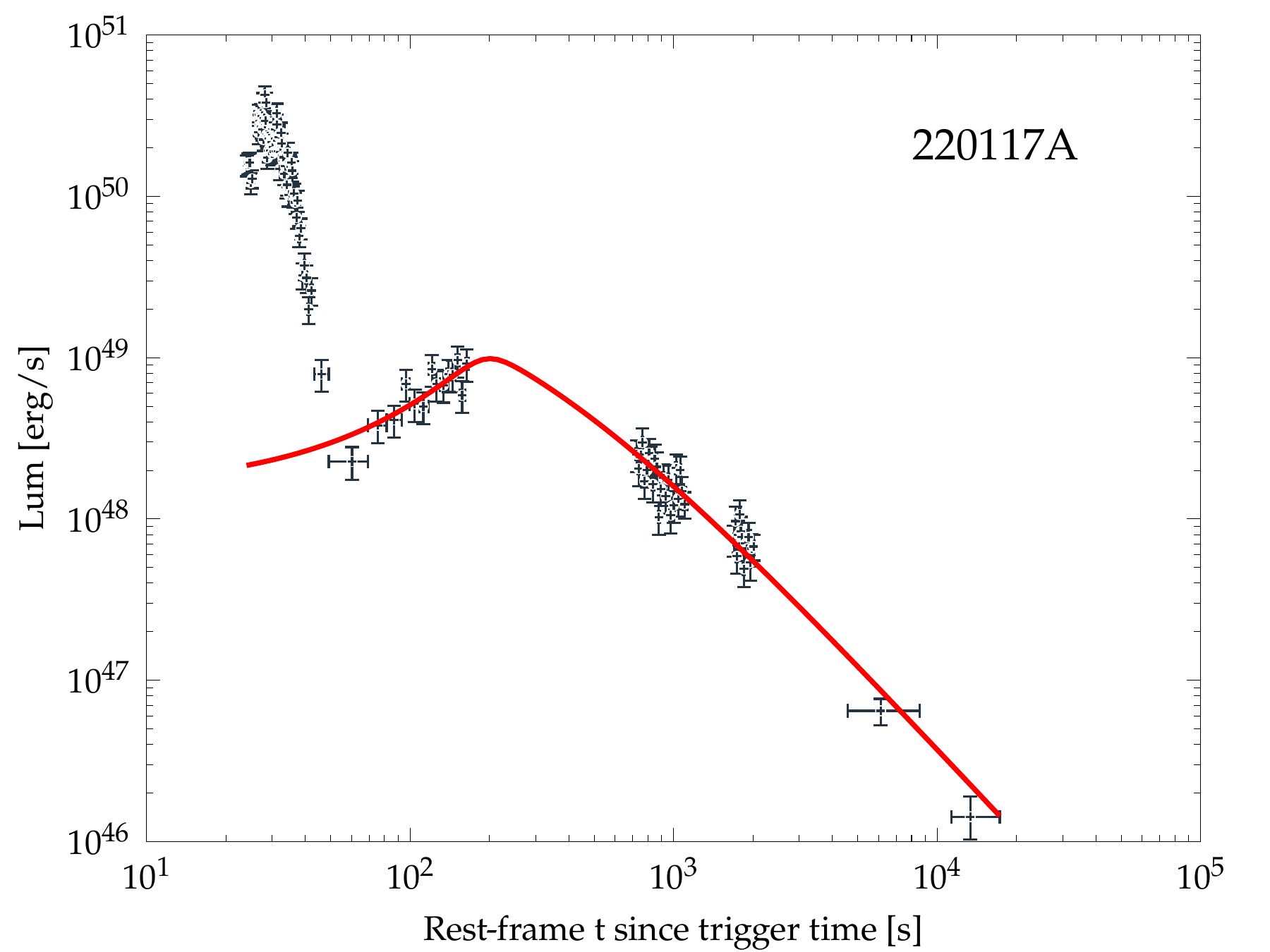}
   \caption{{\it Top}: example of LC successfully modelled with Eq.~\eqref{eq:mod}. {\it Bottom}: example of LC which exhibits a peak following the steep decay and which was modelled with Eq.~\eqref{eq:AGmod} as the afterglow rise caused by the deceleration of the relativistic ejecta. This example is one of the four out of nine GRBs whose rise was succesfully modelled as RS emission.}
   \label{fig:LC_examples}
\end{figure}

The modelling was carried out within a Bayesian approach using the Python {\tt emcee} package (v.3.1.6) \citep{Foreman13}: we adopted the negative log-likelihood of Eq.~\eqref{eq:loglik} with a Monte Carlo Markov Chain (MCMC) 
\begin{equation}
    {\cal L}(L_0, T_0, \alpha)\ =\ \frac{\chi^2}{2}\ =\ \frac{1}{2}\,\sum_{i=1}^{N} \Big(\frac{L_{x,i} - L_x(t_i)}{\sigma_{L_{x,i}}}\Big)^2\;,
    \label{eq:loglik}
\end{equation}
where $L_{x,i}$ and $\sigma_{L_{x,i}}$ are the observed luminosity and its uncertainty at time $t_i$, respectively, and $N$ the number of points. Uniform prior distributions were assumed for $\log{L_0}$ in $[38,58]$, for $\log{T_0}$ in $[0,8]$ and for $\alpha$ in $[0.2,5]$. The corresponding mean values and standard deviations of the marginalised posterior distributions are taken as best-fit values and 1$\sigma$ uncertainties, respectively.
We validated each solution through a $\chi^2$ test, by imposing a minimum threshold of $10^{-3}$ on the p-value. Each solution was inspected visually: in a number of cases, although the overall behaviour of the LC was satisfactorily modelled, the $\chi^2$ test would reject the best-fit model: this was due to the presence of fast and uncorrelated variability modulating the overall profile. We saw now reason to exclude these cases (limited variability modulating the afterglow could result from small inhomogeneities in the shocked medium, or alternatively, be the result of underestimated uncertainties on the flux). We therefore decided to systematically apply the runs test\footnote{We used {\tt runstest\_1samp} to the residuals from the Python module {\tt statsmodels.sandbox.stats.runs}.} to ensure that no trend in the residuals (due to poor modelling) was present. Finally, a solution was accepted if the p-value was $\ge 10^{-3}$ for at least one of the two tests ($\chi^2$ and runs). The top panel of Figure~\ref{fig:LC_examples} shows the case of long GRB\,091018 as an example of successful modelling: this GRB is representative of the quality of the data set, given its rest-frame time coverage and number of points (127); its best-fit parameters and errors are also typical ($\log{L_0}=48.52\pm0.02$, $\log{T_0}=2.24\pm0.03$, $\alpha=1.27\pm0.02$).

Unlike \citetalias{Liang09}, we did not distinguish the cases that show a well defined plateau from the simple PL ones: for the latter cases, we ended up with unconstrained $T_0$ and a 3$\sigma$ upper limit was calculated. The line of reasoning behind our approach is that $T_0$ is likely continuously distributed, therefore the modelling with Eq.~\eqref{eq:mod} should be carried out on equal footing, encompassing the plateau clear-cut cases (for large values of $T_0$) all the way down to the cases which just show a gentle steepening or no slope change altogether (for relatively small values of $T_0$).

As a result, 376 L-GRBs, 32 S-GRBs, and 16 SEE-GRBs passed the selection. Initially, there were ten L-GRBs which did not pass, because their LCs clearly show a rebrightening that follows the steep decay. For these cases, we devised a different modelling, which is presented below.

In the prior activity model, this may be the case when the prompt gamma-ray emission ends before the deceleration of ejecta that were emitted at the zero time. Consequently, instead of adopting Eq.~\eqref{eq:mod}, we modelled the rise and following decay as it is predicted for a homogeneous medium in two alternative cases, depending on which shock (reverse or forward) is the dominant component. Each of these LCs was modelled as
\begin{equation}
    L_x(t)\ =\ L_0\ \Bigg[ \frac{1 - \alpha_1/\alpha_2}{\Big(\frac{t+T_0}{t_{\rm p}}\Big)^{n\alpha_1} +
\Big(\frac{t+T_0}{t_{\rm p}}\Big)^{n\alpha_2} \Big(-\frac{\alpha_1}{\alpha_2}\Big)} \Bigg]^{1/n}
    \label{eq:AGmod}
\end{equation}
where $\alpha_1<0$ and $\alpha_2>0$ are the rise and decay PL indices, $t_p$ the peak time, $L_0$ the luminosity at peak, and the smoothness parameter $n$ was fixed to 5 (from Eq.~1 of \citealt{Guidorzi14}). Instead of treating $\alpha_1$ and $\alpha_2$ as free parameters, we expressed them as a function of the PL index $p$ which describes the energy distribution of the shock-accelerated electrons that emit the afterglow synchrotron radiation. In particular, assuming the slow cooling regime and the synchrotron frequency $\nu_m$ lying below X-rays, for the FS case it is $\alpha_1 = -3$ and $\alpha_2 = 3(p-1)/4$, whereas for the RS it is $\alpha_1 = (3 - 6p)/2$ and $\alpha_2=(3p+1)/4$ \citep{Kobayashi00a,Gao13}. Compared with Eq.~\eqref{eq:mod}, there are two additional parameters: $t_p$, measured since the true zero time, and $p$, whereas $\alpha$ is no more present.
We assumed a uniform prior distribution for $p$ in the range $[2,8]$, while $t_p$ shared the same prior of $T_0$, with the constraint $t_p>T_0$ (otherwise no afterglow peak would be visible in the data).
Like for the majority of cases that were modelled with Eq.~\eqref{eq:mod}, we rejected the solutions with p-values $<10^{-3}$ for both $\chi^2$ and runs test. The bottom panel of Figure~\ref{fig:LC_examples} shows an example of successful modelling with RS. As a result, only 1/10 GRBs was discarded. Hereafter, the 9 successful GRBs are referred to as the AG-rise sample.

%%%%%%%%%%%%%%%%%%%%%%%%%%%%%%%%%%%%%%%%%%%%%%%%%%%%
\section{Results}
\label{sec:results}
%%%%%%%%%%%%%%%%%%%%%%%%%%%%%%%%%%%%%%%%%%%%%%%%%%%%
We restricted the following statistical analysis to the GRBs with well constrained parameters: for a GRB to be considered, its uncertainties had to fulfil the following conditions: $\delta(\log{L_0})\le0.3$, $\delta(\log{T_0})\le0.3$, $\delta\alpha\le0.5$. Similarly, for the AG-rise sample we also imposed $\delta(\log{t_p})\le0.3$, $\delta p\le 0.2$.
Consequently, our sample shrank to 273 L-GRBs (9 of which belonging to the AG-rise set), 19 S-GRBs, and 8 SEE-GRBs, for a total of 300~GRBs with acceptable solutions and well-constrained parameters.

In agreement with the results obtained by \citetalias{Liang09} on a smaller sample, $T_0$ is not correlated with the prompt gamma-ray emission duration, $T_{90}$. We found no correlation between $T_0$ and the intrinsic column density $N_{H,z}$, which is the rest-frame hydrogen-equivalent column density in excess of the Galactic value for any given GRB direction, which is responsible for the photoelectric absorption of soft X-rays.\footnote{This was derived through modelling the late time XRT spectrum available at the Leicester website with the {\sc xspec} model {\tt cflux*TBabs*zTBabs*powerlaw}, where for the Galactic $N_H$ the estimate from \citet{Willingale13} was adopted.}

\begin{figure}[!h]
%\centering
   \includegraphics[width=0.45\textwidth]{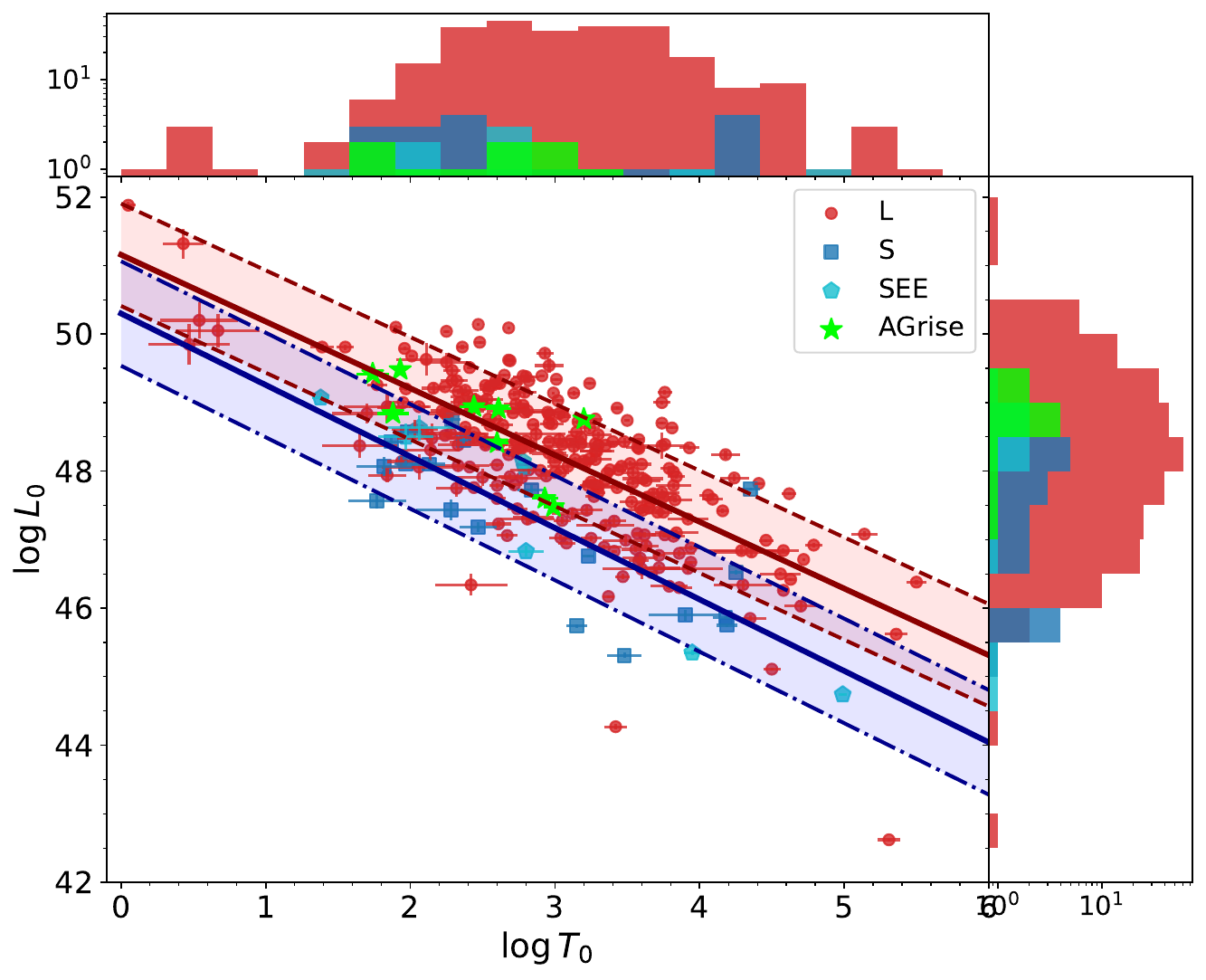}
   \includegraphics[width=0.45\textwidth]{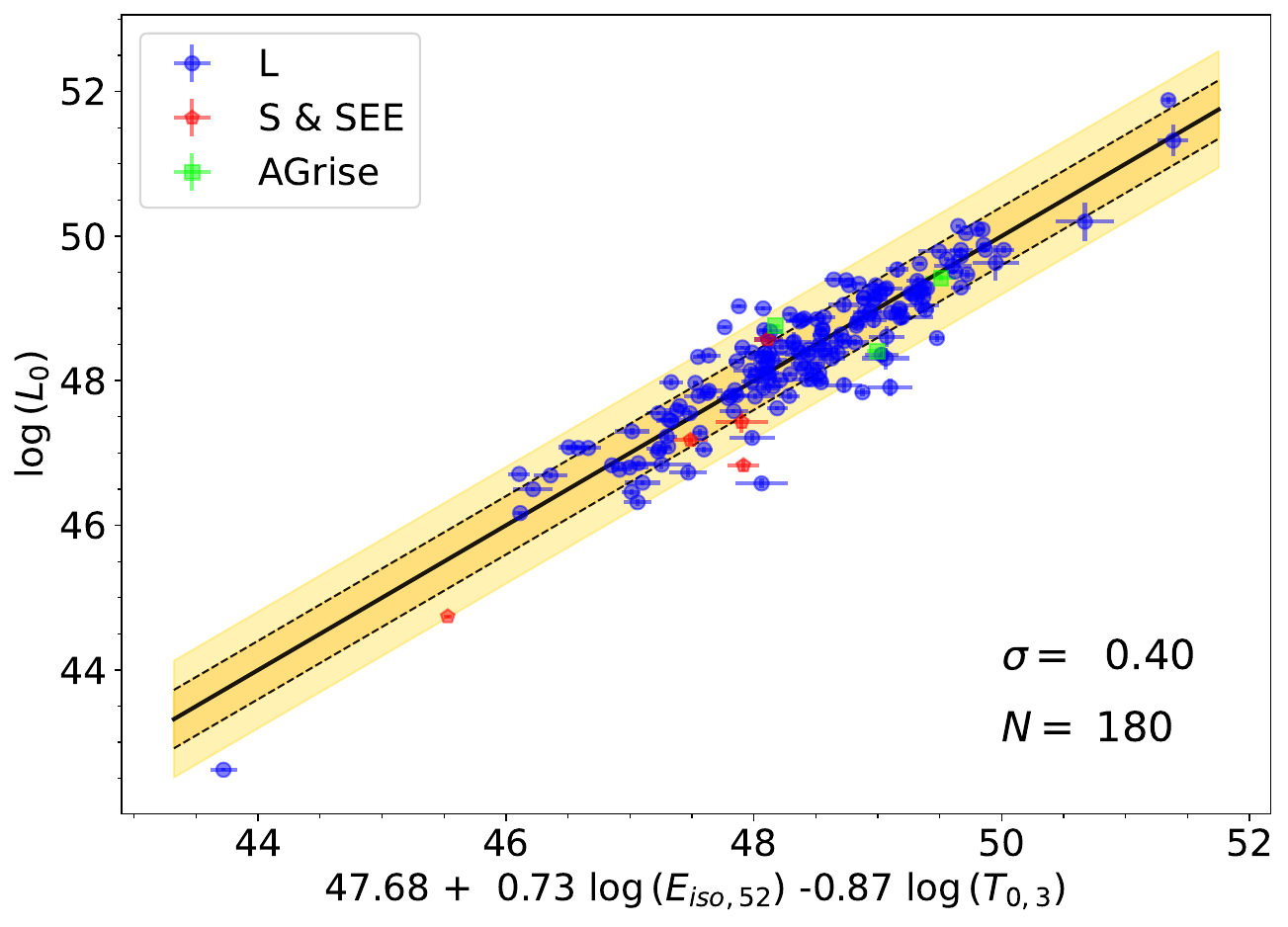}
   \caption{{\it Top}: $L_0$ vs. $T_0$ for two separate groups: (i) L-GRBs (red circles, excluding the AG-rise cases), whose best-fit PL and 1$\sigma$ uncertainty region are shown by the red solid line and red-shaded area, respectively; (ii) S+SEE-GRBs, blue squares and cyan pentagons, respectively, whose best-fit PL is shown with the blue solid line within the blue shaded area of the 1$\sigma$ region. The L-GRBs of the AG-rise sample (green stars) were not used for the modelling. Also shown are the corresponding marginalised distributions. {\it Bottom}: $L_0$-$T_0$-$E_{\gamma,{\rm iso}}$ relation for the 180~GRBs with accurate measurements available, along with the best-fit model and 2$\sigma$ uncertainty region.}
   \label{fig:L0_vs_T0}
\end{figure}

Conversely, $T_0$ is inversely proportional to $L_0$, although with significant scatter. Calling $y=\log{L_0}$ and $x=\log{T_0}$, we modelled $y = m x + q$ adopting the \citet{DAgostini05} likelihood, where the free parameters are $m$, $q$, and the intrinsic scatter $\sigma$.
We modelled L-GRBs (excluding the 9 AG-rise cases) separately from the S-GRBs and SEE-GRBs, which were merged. Figure~\ref{fig:L0_vs_T0} shows the results and Table~\ref{tab:dago2} reports the best-fit parameters.
\begin{table}
\centering
\caption{Best-fit parameters obtained with the \citet{DAgostini05} method for the $L_0$-$T_0$ PL relation for two separate groups: L-GRBs and S-GRBs merged with SEE-GRBs (top panel of Fig.~\ref{fig:L0_vs_T0}).}
\label{tab:dago2}
\begin{tabular}{crccc}
\hline
Sample &  $N$ & $m$    & $q$ & $\sigma$\\
\hline
L & 273 & $-0.97\pm 0.06$ & $51.16\pm 0.18$ & $0.75\pm0.03$\\
S + SEE & 27 & $-1.04\pm 0.16$ & $50.30\pm 0.48$ & $0.76\pm0.12$\\
\hline
\end{tabular}
%\begin{list}{}{}
%\item[$^{\mathrm{(a)}}$]{Number of GRBs with 90\%-significant %estimates of $V_f$ that were used in each case.}
%\end{list}
\end{table}
For both groups the PL index $m$ is compatible with $-1$, that is the inverse proportionality between $L_0$ and $T_0$. They also share the same scatter of $0.75$~dex. Only the normalisation is different, with the mean luminosity of L-GRBs being almost 10 times higher for a given $T_0$. This relation shows that the difference between very luminous afterglows that decay with a simple PL with no signs for plateau, such as 061007 \citep{Mundell07}, and canonical X-ray afterglows may be only apparent and simply reflect the different delays between the true start time of the catastrophic event and the observed prompt gamma-ray emission.

We further explored the possible connection of the X-ray afterglow parameters extracted through the present analysis with other key observables that characterise the prompt gamma-ray emission: the isotropic-equivalent released energy $E_{\gamma,{\rm iso}}$, the peak luminosity $L_{\gamma,{\rm iso}}$ (both referred to the 1--$10^4$~keV rest-frame band), and the intrinsic peak energy of the time-average $\nu\,F_\nu$ spectrum, $E_{\rm p,i}$. We searched preferentially through GRB catalogues with broad passband, which ensures a reliable spectral modelling. In particular, values were taken from either the Konus-WIND catalogues \citep{KWGRBcat17,KWGRBcat21} or elsewhere for some recent GRBs (e.g., \citealt{Guidorzi24b}). We ended up selecting 175 L-GRBs (3 of which are AG-rise), 3 S-GRBs, and 2 SEE-SGRBs, for a total of 180 GRBs.
We searched for possible correlations applying the principal component analysis (PCA)\footnote{We used the Python {\tt sklearn} package (v.1.4.2).}. As a result, we found a three-parameter relation between $L_0$, $T_0$, and $E_{\gamma,{\rm iso}}$ which is common to all GRB classes considered in this work and whose scatter, $0.4$~dex, is significantly smaller than that of the $L_0$-$T_0$ relation. Adopting the \citet{DAgostini05} likelihood, the best-fit relation is given by,
\begin{eqnarray}
   \label{eq:L0_T0_Eiso}
   \log{L_0} & = & (47.68\pm0.04)\ +\ (0.73\pm0.04)\,\log{E_{{\gamma,{\rm iso}},52}}\ + \\\nonumber
   & & - (0.87\pm0.04)\,\log{T_{0,3}}\;
\end{eqnarray}
with an intrinsic dispersion $\sigma=0.404\pm0.024$, where $E_{{\gamma,{\rm iso}},52} = E_{\gamma,{\rm iso}}/(10^{52}~{\rm erg})$, $T_{0,3}=T_0/10^3$ (bottom panel of Fig.~\ref{fig:L0_vs_T0}).
A very similar result was obtained with $L_{\gamma,{\rm iso}}$ replacing $E_{\gamma,{\rm iso}}$, with an intrinsic scatter slightly larger, $\sigma=0.426\pm0.025$, although compatible within errors. The analogous relation of Eq.~\eqref{eq:L0_T0_Eiso} is,
\begin{eqnarray}
   \label{eq:L0_T0_Liso}
   \log{L_0} & = & (48.01\pm0.04)\ +\ (0.75\pm0.04)\,\log{L_{{\gamma,{\rm iso}},52}}\ + \\\nonumber
   & & - (0.69\pm0.05)\,\log{T_{0,3}}\;.
\end{eqnarray}
Finally, we found a 4-parameter correlation with a slightly smaller scatter ($\sigma=0.373\pm0.022$~dex), by adding the PL index $\alpha$ as fourth quantity to the relation of Eq.~\eqref{eq:L0_T0_Eiso},
\begin{eqnarray}
   \label{eq:L0_T0_Liso_a}
   \log{L_0} & = & (47.15\pm0.04)\ +\ (0.67\pm0.04)\,\log{E_{{\gamma,{\rm iso}},52}}\ + \\\nonumber
   & & - (0.97\pm0.05)\,\log{T_{0,3}}\ +\ (0.37\pm0.08)\,\alpha\;.
\end{eqnarray}
Apart from Eq.~\eqref{eq:L0_T0_Liso_a}, the PL index $\alpha$ is not correlated with any individual quantity. Its mean, median, standard deviation for the L-GRBs are $1.61$, $1.44$, and $0.62$, respectively. The analogous values for the S- and SEE-GRBs group are $2.2$, $2.1$, $1.1$; a Kolmogorov-Smirnov two-population test yields a p-value of $3\times10^{-4}$ that they share a common parent distribution. The same comparison applied to the two sets of $T_0$ cannot reject a common distribution (top panel of Fig.~\ref{fig:L0_vs_T0}). The whole sample of $T_0$ values is log-normally distributed with $\mu(\log{T_0})=3.0$ and $\sigma(\log{T_0})=0.8$. Thus, while the distribution of the delay between zero time and prompt gamma-ray emission is similar for the two groups, on average L-GRB X-ray afterglows decay slightly more shallowly than the S+SEE group.

%%%%%%%%%%%%%%%%%%%%%%%%%%%%%%%%%%%%%%%%%%%%%%%%%%%%
\section{Discussion and conclusions}
\label{sec:disc_conc}
%%%%%%%%%%%%%%%%%%%%%%%%%%%%%%%%%%%%%%%%%%%%%%%%%%%%
The early X-ray emission covered by {\it Swift}/XRT from as early as 60--100~s since the gamma-ray trigger can be generally decomposed in two components: (1) internal dissipation, whose end is signalled by the steep decay, which can be interpreted as high-latitude emission \citep{Lyutikov06,Zhang06,Zhang09e,Hascoet12b,Uhm15,Lin17,Ajello19}, along with the occasional presence of X-ray flares, whose internal origin is supported by a number of common properties with gamma-ray prompt pulses \citep{Margutti10b,Guidorzi15b}; (2) the X-ray afterglow, emitted by the interstellar medium swept up by a FS and also possibly by the relativistic ejecta as they are decelerated by a RS.
In this work we modelled the latter component within the context of the prior activity model, which assumes the zero time to precede the GRB itself by a time $T_0$. Referring to the correct zero time, the afterglow LC is expected to exhibit a PL rise, followed by a PL decay, with no plateau whatsoever.

For 424 out of 463 GRBs with measured redshift the X-ray afterglow behaviour is consistently modelled with either Eq.~\eqref{eq:mod} or Eq.~\eqref{eq:mod2} prescribed by this model, which consequently proved successful in $\sim90$\% cases. The failure of the model in the remaining 10\% is likely due to on-going internal activity which hampers the identification of the underlying afterglow, or, in the so-called ``internal plateau'' cases (both long and short), in which the drop following the plateau is too steep to be accommodated as afterglow \citep{Troja07,Rowlinson10}.

The possibility that for most GRBs the prompt gamma-ray emission occurs some time ($\sim10^2$--$10^4$~s in the GRB frame) after the beginning of the event appears to be plausible for a number of reasons: an immediate implication of this model is that one should observe the same behaviour in the X-ray and optical afterglow LCs. First of all, the distribution of the PL index $\alpha$ derived by us is consistent with the typical values observed both in X-rays and in the optical \citep{Zaninoni13}. On the other hand, for a sizeable fraction ($\sim 40$\%; \citealt{Panaitescu11}) of early afterglows, optical and X-rays display different behaviours, especially in the first hundreds seconds \citep{Melandri08,Oates11}: this can be ascribed in most cases to on-going internal activity which affects the X-ray emission and, occasionally, the optical emission too \citep{Kopac13}, along with possible changes in the optical PL decay caused by synchrotron break frequencies crossing the observed passband. Nonetheless, for 60--75\% cases, optical and X-ray show similar behaviours and LC morphologies, including the occurrence of plateaus \citep{Oates09,Panaitescu11,Roming17}. Moreover, the higher fraction of initially rising optical afterglows than in the X-ray band is possibly due to the on-going internal activity that may hide the simultaneous X-ray afterglow rise. In parallel, the numerous optical afterglows that are seen to decay from early on, imply that the ejecta deceleration and consequent afterglow peak must have occurred very early, potentially even before the GRB itself as in the prior activity model.

Our results show that the rest-frame delay, $T_0$, between zero time and the gamma-ray trigger time, where the latter approximately corresponds to the start time of GRB itself, is continuously distributed, with values as small as a few ten seconds or less (Fig.~\ref{fig:L0_vs_T0}). Whenever $T_0$ is a few seconds or less, the deviation from a straight PL becomes almost negligible using the trigger as zero time and in this case no plateau is expected. A nice example of this case is offered by the luminous 061007: its broadband afterglow looks like an unbroken PL from $\sim 2$~minutes from the GRB onset all the way to several days with $\alpha\sim1.7$ \citep{Mundell07}. In our analysis, the X-ray LC shows evidence for a tiny but statistically significant deviation from a straight PL with $\log{T_0} = 0.43\pm0.14$, that is just a few seconds before the gamma-ray trigger time. So, in practice the onset of its prompt gamma-ray emission essentially marks the beginning of the event within a few seconds tolerance. The paucity of small delays in our sample is likely due to the fact that a high S/N is required to accurately measure a small curvature in the PL decay.

Another advantage of interpreting our results within the prior activity model is the common description of two apparently different behaviours: the plateau vs. the rising X-ray afterglow (our AG-rise sample). The only difference between the two behaviours is that for the former class, the afterglow peak must have occurred before the end of the GRB, whereas for the latter group it is the opposite, so that the steep decay flux drops quickly and deeply enough to let the rising afterglow emerge. To test our interpretation of the rising afterglows as due to the ejecta deceleration, for the GRBs of our AG-rise sample, we searched for simultaneous optical observations in the literature: for the 4/9 events with available data, we found an optical peak consistent with ejecta deceleration, as also argued by the various authors: 110213A \citep{Wang22}, 140515A \citep{Melandri15}, 181110A \citep{Han22}, 190829A \citep{Dichiara22,Salafia22}. That for all the testable GRBs of our AG-rise sample the optical data confirm the ejecta deceleration scenario, further corroborates our interpretation of the X-ray afterglow rise.

Thanks to the capability of identifying and localising X-ray transients over a large field of view, recent EP discovered X-ray emission that lasted very long ($\sim 10^3$~s) and preceded (by up to several hundred seconds) the prompt gamma-rays for a number of GRBs that were successively detected by some GRB experiment among {\it Fermi}, {\it SVOM}, {\it Swift} (Table~\ref{tab:EP}). In addition to the discovery of so-called ``ultra-long'' GRBs \citep{Levan14}, these discoveries make the case for GRB engines that can operate much longer than what was previously thought and whose soft emission can start correspondingly earlier than gamma-rays. Although alternative interpretations to a GRB, such as a tidal disruption event, have been put forward, the recent case of the day-long EP250702a/GRB\,250702B\footnote{GRB\,250702B, D, and E are from the same source. However, to avoid confusion, we adhere to the GRB naming convention and refer to it as 250702B \citep{BurnsSvinkin25}.} with the X-ray emission preceding the prompt gamma-rays by one day, could be another compelling example \citep{Levan25}.

\begin{figure}[!h]
%\centering
   \includegraphics[width=0.45\textwidth]{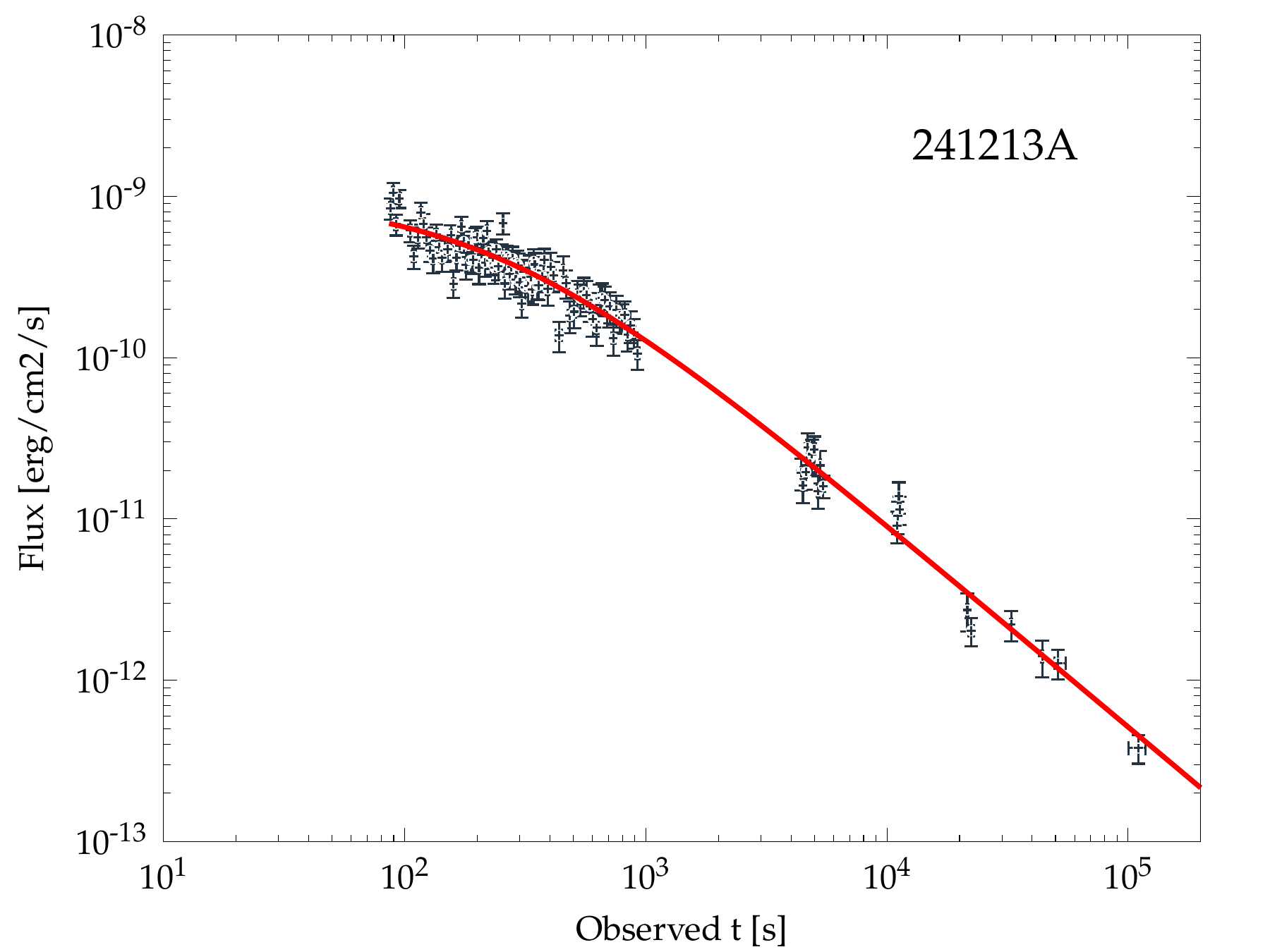}
   \includegraphics[width=0.45\textwidth]{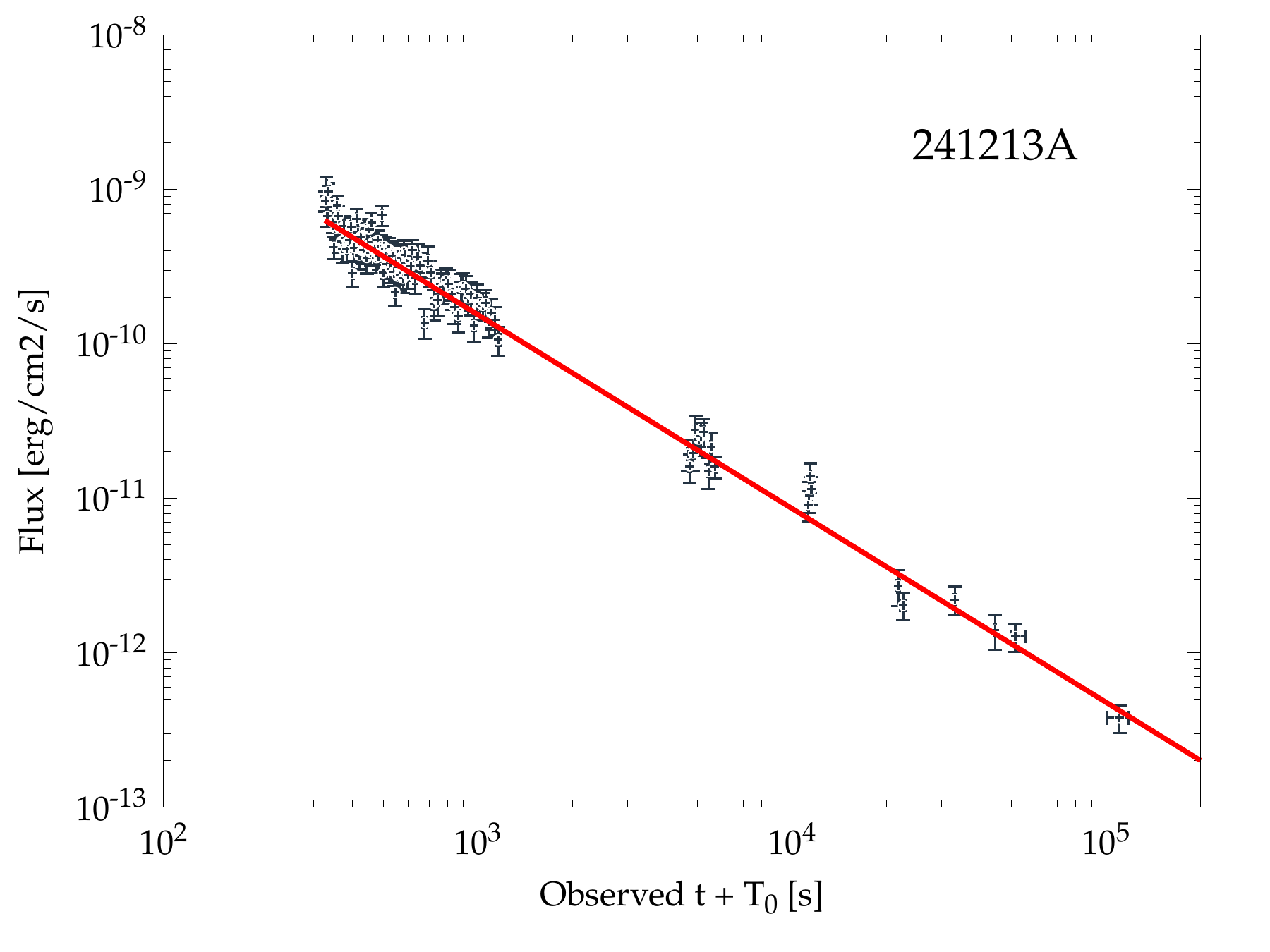}
   \caption{{\it Top}: observed XRT LC of 241213A in the observer frame, which was also detected by {\it Einstein Probe}/WXT 105~s earlier (see Table~\ref{tab:EP}). A successful modelling with a different zero time yields $T_0=250_{-50}^{+70}$~s. {\it Bottom}: same LC as in the top panel, but with the reference time set to 250~s prior to the {\it Swift} trigger time. The LC  is modelled with a SPL. The observed delay between soft X- and gamma-rays is comparable with the value inferred from the prior activity model.}
   \label{fig:241213A}
\end{figure}

\begin{table*}
\centering
\caption{Recent examples of GRBs, whose X-ray emission was detected by {\it Einstein Probe} long before the prompt gamma-ray emission.}
\label{tab:EP}
\begin{tabular}{llccrcc}
\hline
EP name & GRB name & Trigger Time EP & GRB Trigger time & Delay$^{\rm (a)}$ & References$^{\rm (b)}$ & $T_0^{\rm (c)}$\\
     &   &     (UT)        &       (UT)       &  (s)  &  & (s) \\
\hline
EP240315a & 240315C & 20:10:44 & 20:16:58 & 370 & (1) & N/A\\
EP240802a & 240802A & 10:32:52 & 10:34:03 & 110 & (2,3) & N/A\\
EP240913a & 240913C & 11:39:33 & 11:42:36 & 180 & (4,5) & N/A\\
EP241026a & 241026A	& 22:41:28 & 22:42:32 &  60	& (6,7) & N/A \\
EP241213a$^{\rm (d)}$ & 241213A & 02:17:15 & 02:19:00 & 105 & (8,9) & $250_{-50}^{+70}$\\
EP250615a$^{\rm (d)}$ & 250615A	& 22:25:17 & 22:25:20 &   3	& (10,11) & $<13$\\
\hline
\end{tabular}
\begin{list}{}{}
\item[$^{\mathrm{(a)}}$]{Calculated as $t_\gamma-t_X$, where $t_\gamma$ and $t_X$ are the GRB trigger time and the EP X-ray trigger time, respectively.}
\item[$^{\mathrm{(b)}}$]{References: (1) \citet{Liu24_EP240315a}; (2) \citet{EP240802a}; (3) \citet{EP240802a_SVOM}; (4) \citet{EP240913a}; (5) \citet{240913a_GBM}; (6) \citet{EP241026a}; (7) \citet{EP241026a_SVOM}; (8) \citet{EP241213a}; (9) \citet{EP241213a_BAT}; (10) \citet{EP250615a}; (11) \citet{EP250615a_BAT}.}
\item[$^{\mathrm{(c)}}$]{$T_0$ in the observer frame, since no redshift is available. The upper limit is given at 90\% confidence.}
\item[$^{\mathrm{(d)}}$]{Thanks to the prompt follow up with {\it Swift}/XRT, it is possible to model the early X-ray emission and test the prior activity model. That is why we included EP250615a in spite of a small delay between X- and gamma-rays.}
\end{list}
\end{table*}
The data in Table~\ref{tab:EP} were obtained by sifting  the General Coordinates Network (GCN) notices\footnote{\url{https://gcn.nasa.gov/circulars}.} up to July 2025, selecting the common detections by both EP and gamma-ray instruments that routinely monitor the prompt gamma-ray emission.
Table~\ref{tab:EP} reports two EP events that also triggered {\it Swift} and for which XRT recorded a prompt X-ray afterglow LC, which allowed us to test the prior activity model. In particular, the X-ray LC of EP241213a/GRB\,241213A can be successfully modelled, yielding $\log{T_0} = 2.4\pm 0.1$ (observer time), corresponding to $250_{-50}^{+70}$~s (top panel of Fig.~\ref{fig:241213A}). EP/Wide-Field X-ray Telescope (WXT) detected X-ray emission 105~s before, so not as early as our result, but comparably so (and in any case an even earlier start than the EP/WXT time is possible in principle). The bottom panel of Fig.~\ref{fig:241213A} shows the same LC with the zero time obtained by us, which appears as a SPL. One concludes that the indication supplied by the modelling of the early X-ray LC gives a hint about the true start of the phenomenon, which soft X-ray observations seem to support.

The case of EP250615a/GRB\,250615A represents an interesting and complementary example: the XRT light curve is a SPL from $100$~s all the way to at least $10^4$~s, with $\alpha=1.6$\footnote{\url{https://www.swift.ac.uk/xrt_live_cat/01324646/}.}. This inevitably implies a small delay between X- and gamma-rays and indeed the X-ray detection by EP/WXT differs by just a few seconds (Table~\ref{tab:EP}), which is consistent with the 90\% confidence upper limit of 13~s on $T_0$ (observer frame) derived by us.

The case of the brightest GRB yet recorded, GRB\,221009A, deserves a specific comment. At first glance, the XRT light curve looks like a SPL from $10^3$ to $10^7$~s (rest). However, following \citet{Williams23} a broken PL provides a better description, with a PL index changing from $1.50$ to $1.67$ around a rest-frame break time of $7\times10^4$~s, although the exceptional statistical quality of the data shows evidence for a more complex behaviour. In our analysis, GRB\,221009A did not make it to the selected sample, due to the very poor modelling evidenced by the p-values of both $\chi^2$ and runs tests ($7\times10^{-31}$ and $6\times10^{-16}$). The complexity of this behaviour, foregrounded by the unprecedented statistical quality of the data set, might arise from the presence of residual internal activity. 
Observations in the TeV range by the Large High Altitude Air Shower Observatory (LHAASO) identified the afterglow reference time at $T_* = 226$~s (observed) since the {\it Fermi}/Gamma-ray Burst Monitor (GBM) trigger time \citep{Lesage23}, corresponding to rest-frame 196~s \citep{LHAASO23_sci}. This implies that the afterglow would have begun right at the start of the pinnacle of the keV-MeV activity in the observed $220-280$~s window \citep{Frederiks23,Lesage23}. Adopting $T_*$ as reference time, \citet{LHAASO23_sci} found evidence for a jet break at $T_*+670$~s, which corresponds to a rest-frame time of $(T_*+670)/(1+z)=780$~s since the prompt keV-MeV emission onset. The XRT light curve begins after the putative jet break time, thus undermining the applicability of our analysis for this special GRB, as also evidenced by the poor modelling itself.
Nonetheless, one cannot help but notice that GRB\,221009A, like other extremely bright GRBs, shows no canonical behaviour in the early X-ray emission, as also noticed by \citet{OConnor23}. TeV observations provide compelling evidence that the afterglow reference time occurs very close to the peak of the prompt gamma-ray emission. This automatically implies that no canonical X-ray plateau should be observed, as indeed is the case, in agreement with the basic prediction of the prior activity model.

\begin{figure}[!h]
   \includegraphics[width=0.48\textwidth]{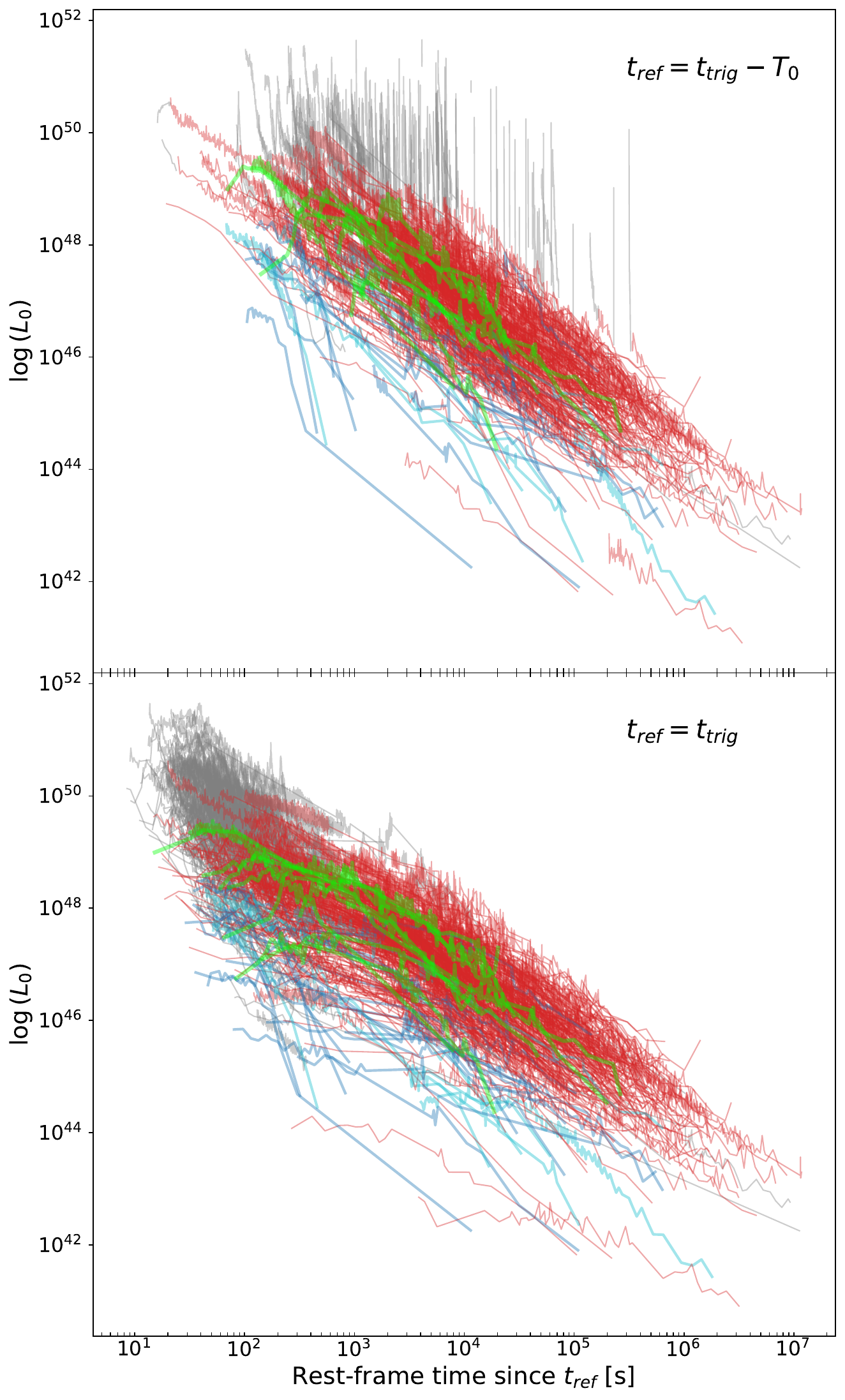}
   \caption{{\it Top}: all the LCs displayed by shifting the rest-frame zero time backward by $T_0$. Red, blue, cyan, and green lines correspond to L-GRBs, S-GRBs, SEE-GRBs, and AG-rise cases, respectively. The grey portions were ignored by the modelling, being interpreted as internal activity. {\it Bottom}: same data as in the top panel, except that the reference time coincides with the BAT trigger time.}
   \label{fig:Alle_zusammen}
\end{figure}

There are several implications of the results: (a) the inferred gamma-ray efficiency decreases, since the energy released in the plateau is no more internal dissipation \citep{Ioka06}; (b) the small fraction of long AG-rise cases (10/376) suggests that in most cases the deceleration occurs before the prompt gamma-ray emission or, at least, before its end. Within the thin shell approximation, the deceleration time $t_p$ through a homogeneous medium with particle density $n$ is
\begin{equation}
    t_p\ \simeq\ \Big(\frac{3\,E_{\rm iso}}{32\pi \Gamma_0^8 n m_p c^5} \Big)^{1/3}\ \simeq\ (90~{\rm s})\ E_{{\rm iso},52}^{1/3}\,\Gamma_{0,2}^{-8/3}\,n^{-1/3}\;,
    \label{eq:tp}
\end{equation}
where $E_{{\rm iso},52} = E_{\rm iso}/10^{52}$~erg is the released kinetic energy, $\Gamma_{0,2}=\Gamma_0/10^2$ is the initial Lorentz factor \citep{Sari99}. Assuming typical values for L-GRBs, Equation~\eqref{eq:tp} yields $t_p$ values that are shorter than most of our $T_0$ estimates, thus explaining the small fraction of AG-rise cases. Another implication (c) concerns the jet break times, which would become longer by $T_0$ than currently estimated, thus implying wider jet angles. Lastly, (d) GRB inner engines operate much longer and more frequently than what it has been assumed so far. Figure~\ref{fig:Alle_zusammen} shows the ensemble of the 300 LCs with well constrained parameters both referred to the BAT trigger time (bottom) and to the estimated zero time (top). While the former case shows the flattening caused by most plateaus, the average behaviour of the latter is a SPL with a notable on-going internal activity (grey traits) superposed to the PL decay and extending all the way up to $10^5$~s.

Our analysis reveals a significant delay between the onset of the explosion and the prompt gamma-ray emission in short GRBs as well. The relativistic outflows appear to begin approximately 30 to $10^4$ seconds before the gamma-ray trigger (see Fig.~\ref{fig:L0_vs_T0}). These delays are substantially longer than the 1.7-second interval observed between the gravitational wave signal GW170817 and the short GRB\,170817A \citep{LIGO-Fermi17}.

In the case of GRB 170817A, the observed off-axis jet means the early X-ray afterglow does not reliably constrain the explosion time ($T_0$). However, the gravitational wave signal provides a precise timestamp for the neutron star merger. It is unlikely that a powerful blast wave could be launched before the merger, although mechanisms such as resonant shattering of neutron star crusts may release energy on the order of $\sim 10^{47}$ erg, a few orders of magnitude lower than that of typical blast waves \citep{Neill22}. 
Consequently, the explosion time offset for GRB 170817A is expected to be less than 1.7 seconds. As illustrated by the EP events (long-duration cases) listed in Table~\ref{tab:EP}, the value of $T_0$ may vary significantly from event to event. 

While extended emission in short bursts and recently observed long-duration merger events suggest that compact object mergers can power jets for durations exceeding 100 seconds, sustaining energy output for up to $10^4$ seconds would demand a revised understanding of the central engine.

Numerous past investigations modelled the canonical X-ray afterglows through the end-of-plateau time, $T_a$, along with the X-ray luminosity at that time, $L_a$. It was first discovered an inverse proportionality between $T_a$ and $L_a$ for 77~GRBs \citep{Dainotti10}. Using 55 optimally selected GRBs, \citet{XuHuang12} then found the three-parameter correlation $L_a\propto E_{\gamma,{\rm iso}}^{0.88}\,T_a^{-0.87}$ with an intrinsic scatter of $0.43\pm0.05$~dex, and a similar result, with $L_{\gamma,{\rm iso}}$ replacing $E_{\gamma,{\rm iso}}$, was later reported \citep{Dainotti16}.
More recently, the three-parameter correlation was confirmed over a sample of 174 GRBs and refined as 
$L_a\propto E_{\gamma,{\rm iso}}^{0.84\pm0.04}\,T_a^{-1.01\pm0.05}$ with $\sigma=0.39\pm0.03$~dex intrinsic scatter and, remarkably, the same relation was found to hold true for all kinds of GRBs \citep{Tang19}.

That both the $L_a-T_a$ inverse proportionality and the $L_a-E_{\gamma,{\rm iso}}-T_a$ correlation are very similar to our results (Table~\ref{tab:dago2} and Eq.~\eqref{eq:L0_T0_Eiso}), is not completely unexpected: by construction, $T_0$ is strongly correlated with $T_a$ (at least for the cases for which a plateau can be confidently identified), as well as $L_0$ with $L_a$. Rather, the two sets of parameters do not differ simply by definition, but especially for their meaning and implications, some of which are already discussed above. Moreover, our description makes a further step, by including afterglow rising events, which are also modelled consistently within the prior activity model with similar results to other L-GRBs. Lastly, the different definitions of the two sets of parameters allowed us to extend the three-parameter correlation over nine decades along $L_0$ against the six ones along $L_a$ (see Fig.~8 of \citealt{Tang19}).

The $L_0$-$T_0$ inverse proportionality can be interpreted within the prior activity model as follows: on average, the later the GRB prompt emission occurs since the zero time, the older the X-ray afterglow, that will have faded as a PL with an average index of $-1$, which is a typical value especially from $10^2$ to $10^4$~s \citep{Oates11} and consistent with the prediction of a FS in slow cooling regime. Conversely, interpreting the $L_0-E_{\gamma,{\rm iso}}-T_0$ correlation is less obvious. Taking $L_0\,T_0$ as a rough proxy of the radiated energy by the early afterglow at the time of the prompt gamma-ray emission, this quantity should be linked to the gamma-ray released energy. A possible way is that the more energetic the first ejecta which are responsible for the afterglow, the more energy is then dissipated into gamma-rays, possibly constraining the way the inner engine releases its energy over time. The reason why short and long GRBs share the same relation is even less obvious, but its validity is an established property that should be considered.

\begin{acknowledgements}
We thank the anonymous reviewer for their helpful comments. This work made use of data supplied by the UK Swift Science Data Centre at the University of Leicester. M.M. and R.M. acknowledge the University of Ferrara for the financial support of their PhD scholarships. M. B. acknowledges the Department of Physics and Earth Science of the University of Ferrara for the financial support through the FIRD 2024 grant.
\end{acknowledgements}

\end{document}